\documentstyle[11pt,newpasp]{article}
\begin{document}

\title{Molecular Gas in High-Redshift Submillimeter Galaxies}
\author{D. T. Frayer and N. Z. Scoville}
\affil{California Institute of Technology, Astronomy Dept. 105-24,
Pasadena, CA 91125, USA}

\begin{abstract}

We present observations of the luminous population of high-redshift
sub-mm galaxies taken at the OVRO Millimeter Array.  Studies of sub-mm
galaxies are vital to our understanding of the formation and early
evolution of galaxies since this population could account for a
significant fraction of the total amount of star formation and AGN
activity at high redshift.  We discuss the CO detections for
SMM\,J02399-0136 at $z=2.8$ and SMM\,J14011+0252 at $z=2.6$.  The CO
data show the presence of massive molecular gas reservoirs ($M({\rm
H}_2)\sim {\rm few}\times 10^{10}$--$10^{11} M_{\odot}$) and suggest
that the sub-mm galaxies are similar to low-redshift, gas-rich
ultraluminous infrared galaxies (ULIGs).  These results highlight the
importance that future mm/sub-mm interferometric observations will have
on our understanding of the high redshift universe.
\end{abstract}


\section{Introduction}

Deep surveys of the submillimeter sky using SCUBA on the JCMT have
uncovered a population of luminous dusty galaxies at high-redshift
(Smail, Ivison \& Blain 1997; Barger et al.  1998; Hughes et al. 1998;
Eales et al. 1999).  Although their redshift distribution is still
uncertain, the majority of the sub-mm galaxies are believed to be at
high redshifts
($z\mathrel{\hbox{\rlap{\hbox{\lower4pt\hbox{$\sim$}}}\hbox{$>$}}} 2$)
based on their radio and near-infrared (NIR) data (Smail et al. 1999a,b;
Carilli \& Yun 1999).  This contrasts somewhat with the early optical
studies which argued for lower redshifts for the sub-mm population
(Lilly et al. 1999; Barger et al. 1999).  Several sub-mm sources are now
thought to be undetected at optical wavelengths which makes radio, mm,
and NIR follow-up studies crucial, and supports the high-redshift
scenario for the sub-mm population.

The relative importance of AGN and starburst activity in powering the
high luminosities of the sub-mm population is still open to question.
If dominated by AGN activity, the sub-mm galaxies would be responsible
for a majority of the X-ray background (Almaini et al. 1999).
Alternatively, if the sub-mm galaxies are predominantly powered by
starbursts, they would contribute significantly to the total amount of
star formation at high redshift (Blain et al. 1999).  If the sub-mm
galaxies have a significant starburst component, we would expect to
detect massive molecular gas reservoirs associated with the
star-formation activity.

At the OVRO Millimeter Array, we have been carrying out CO observations
of the sub-mm population of galaxies in order to constrain their
molecular gas masses, and hence their star-formation activity.  We have
concentrated on the sub-mm galaxies discovered during a survey toward
rich, lensing clusters (Smail et al. 1998).  This sub-mm survey has the
advantage of having the several candidate counter-parts with accurate
redshifts, and the sources are typically brighter than those found in
other surveys due to gravitational amplification by factors of 2--3 from
the foreground cluster.  In addition, the observed flux ratios at
different wavelengths should generally represent their intrinsic values
since weak gravitational lensing from cluster potentials is independent
of the size of the emission regions.

\begin{figure}[t]
\plotfiddle{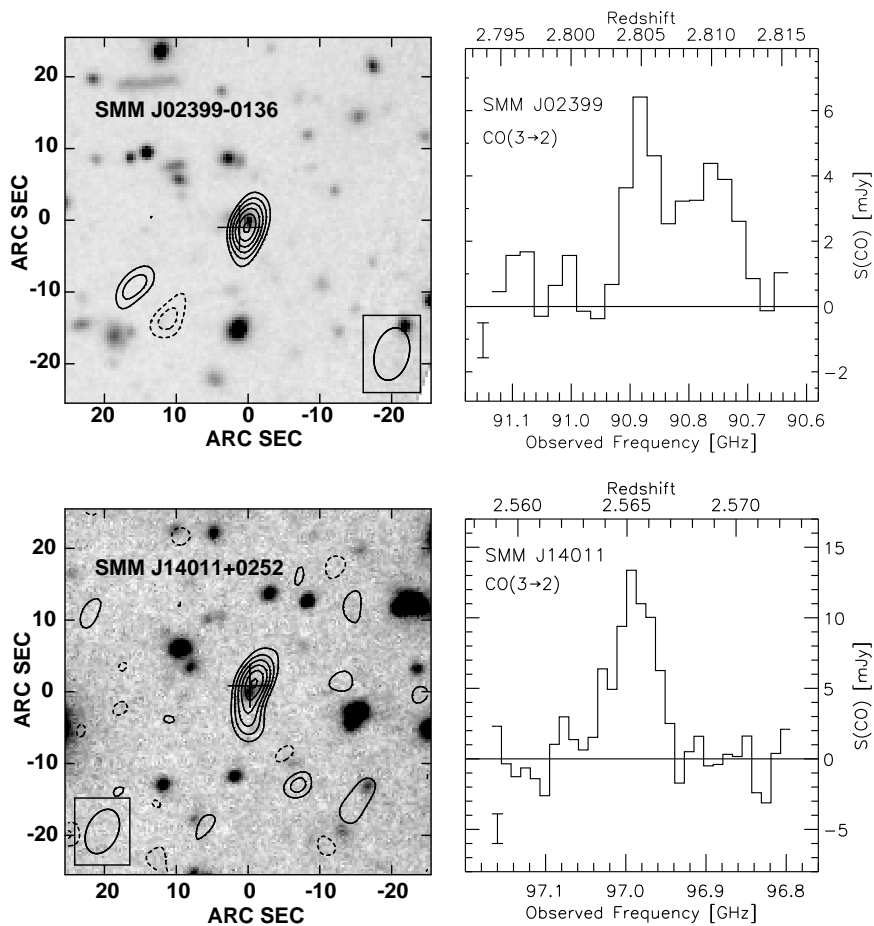}{4.2in}{0.}{65.}{65.}{-210.}{-98.}
\caption{OVRO CO(3-2) detections for SMM\,J02399$-$0136 (top) and
SMM\,J14011+0252 (bottom).  The grey-scale images at the left are optical
images while the contours represent the integrated CO maps for the
sub-mm galaxies.  The CO emission is unresolved for both galaxies.  The
crosses represent the positions of the SCUBA detection.  The
corresponding CO(3-2) spectra are shown at the right(Frayer et al. 1998,
1999).}
\end{figure}

\section{Results and Discussion}

We have detected CO emission from two sub-mm systems (Fig. 1).  The CO
emission is coincident in both position and redshift with their optical
counter-parts, hence confirming their association with the sub-mm
sources.  The first source, SMM\,J02399$-$0136 at $z=2.8$ (SMM\,J02399),
shows an AGN component in its optical spectrum (Ivison et al. 1998),
while the second sub-mm galaxy, SMM\,J14011+0252 at $z=2.6$
(SMM\,J14011), shows only evidence for starburst activity at optical/NIR
wavelengths (Ivison et al. 2000).  Although the optical characteristics
of these galaxies appear vastly different, their radio, sub-mm, and CO
properties are fairly similar and are consistent with a high level of
star formation activity (SFRs of a few$\times
10^{2}$\,M$_{\odot}$\,yr$^{-1}$ to more than
$10^{3}$\,M$_{\odot}$\,yr$^{-1}$, depending on the IMF and AGN
contamination).  After correcting for lensing, we find CO luminosities
of 3--4$\times 10^{10}$\,K\,km\,s$^{-1}$\,pc$^2$
(H$_o=50$\,km\,s$^{-1}$\,Mpc$^{-1}$; $q_o=1/2$) in these two systems.
These CO luminosities correspond to molecular gas masses of about
$5\times 10^{10}$---$2\times10^{11}$\,M$_{\odot}$, depending on the
exact value of the CO to H$_2$ conversion factor (e.g., Solomon et
al. 1997).  Both SMM\,J02399 and SMM\,J14011 appear to be associated
with a merger event.  Given that mergers of gas-rich galaxies at
low-redshift result in massive starbursts, we expect star-formation to
be an important component for powering the far-infrared luminosities in
both of these systems.  In fact, the large molecular gas masses of
SMM\,J02399 and SMM\,J14011 are sufficient to form the stars of an
entire L$^{*}$ galaxy, which suggests that the sub-mm population may
represent the formative phase of massive galaxies.

\begin{figure}[t]
\plotfiddle{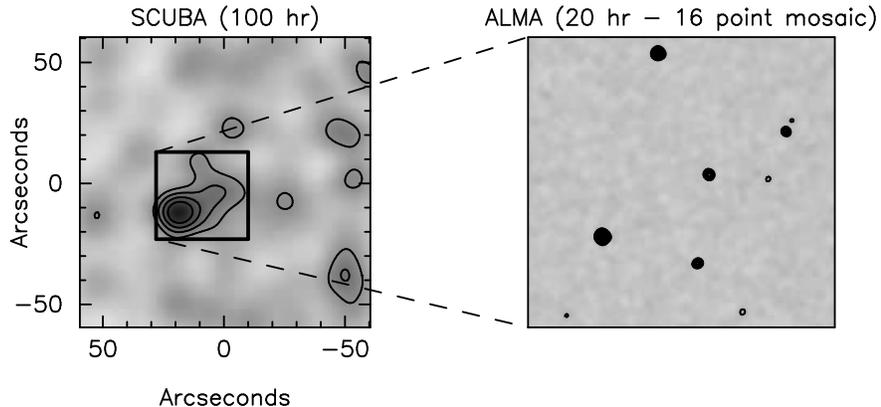}{2.in}{-90.}{65.}{65.}{-270.}{+300.}
\caption{A randomly generated patch of the sky assuming the sub-mm
counts given by Blain et al. (1999).  The simulated SCUBA map (left) is
based on an 100 hour jiggle-map at 850$\mu$m.  The noise level is 0.64
mJy/beam.  The corresponding ALMA map is shown at the right.  The ALMA
predictions assume a 20 hour (including overhead), 16-point mosaic
centered around the brightest SCUBA source.  The rms level is 0.08
mJy/beam, and the contour levels in both maps start at $5\sigma$.  These
results demonstrate the improvement that ALMA will provide over current
observations.}
\end{figure}

\section{Conclusions}

SMM\,J02399 and SMM\,J14011 share many of the same properties of the
local population of ULIGs, such as high infrared (IR) luminosities,
associated with mergers, massive molecular gas reservoirs, comparable CO
line widths, and similar IR/radio and IR/CO luminosity ratios.  Given
that low-redshift ULIGs tend to be comprised of massive starbursts with
varying levels of AGN contamination (Sanders \& Mirabel 1996; Genzel et
al. 1998), we could expect to find similar results for the high-redshift
sub-mm galaxies.  If this is correct and since the sub-mm galaxies are
so numerous (factor of $10^{2}$--$10^{3}$ times more numerous per
comoving volume than low-z ULIGs), the sub-mm population is expected to
contribute significantly
($\mathrel{\hbox{\rlap{\hbox{\lower4pt\hbox{$\sim$}}}\hbox{$>$}}}30$\%)
to both the total amount of star-formation and AGN activity at high
redshift.  Future studies of this population with ALMA will help
elucidate our general understanding of the formation and evolution of
galaxies.

\acknowledgements

The observations presented here were done in collaboration with
A. Evans, M. Yun, and SCUBA lens survey team of I. Smail, R. Ivison,
A. Blain, and J.-P. Kneib.  We thank J. Carpenter for assistance in
developing the simulation code.  We appreciated the efforts of our OVRO
colleagues who have helped make these observations a success.  The OVRO
Millimeter Array is a radio telescope facility operated by the
California Institute of Technology and is supported by NSF grant AST
96-13717.

\end{document}